# Multiple Junction Biasing of Superconducting Tunnel Junction Detectors

K. Segall
*Department of Physics and Astronomy, Colgate University, Hamilton, New York, 13346*

J.J. Mazo
*Departamento de Física de la Materia Condensada and ICMA, CSIC-Universidad de Zaragoza, 50009 Zaragoza, Spain*

T.P. Orlando
*Department of Electrical Engineering and Computer Science, Massachusetts Institute of Technology, Cambridge, Massachusetts, 02139*

We describe a new biasing scheme for single photon detectors based on superconducting tunnel junctions. It replaces a single detector junction with a circuit of three junctions and achieves biasing of a detector junction at subgap currents without the use of an external magnetic field. The biasing occurs through the nonlinear interaction of the three junctions, which we demonstrate through numerical simulation. This nonlinear state is numerically stable against external fluctuations and is compatible with high fidelity electrical readout of the photon-induced current. The elimination of the external magnetic field potentially increases the capability of these types of photon detectors and eases constraints involved in the fabrication of large detector arrays.

Over the past two decades, the use of single photon detectors based on Superconducting Tunnel Junctions (STJs) has received considerable attention.[1] In these detectors, a photon with energy larger than the superconducting energy gap is absorbed in an STJ, creating quasiparticle excitations. These quasiparticles can be read out as a current pulse through the STJ. The integrated charge from this pulse can be used as a measure of the photon energy, giving the detectors inherent spectral resolving power. This technology has been used successfully for photons of energy from 1 eV to 10 keV, with a spectral resolution of order 10-15 eV for photons in the 1-10 keV range. This energy resolution is significantly better than what semiconductor detectors can provide. Besides the energy resolution, these detectors also offer single photon efficiency and a large absorption count rate. If two junctions are used with a single absorber, they can offer spatial imaging capabilities with only a few readout channels.[2]

In order to operate properly, the STJ detector must be biased at a voltage between zero and ($2\Delta/e$), where $2\Delta$ is the energy gap of the superconductor and $e$ is the electron charge. This range is known as the subgap region. To bias stably in the subgap region, a small magnetic field is usually applied parallel to the junction in order to suppress the Josephson supercurrent. A stable bias without a magnetic field is theoretically possible, but in practice usually results in significant signal reduction and/or added noise in the photon pulse readout.[3] While the application of a parallel field is not difficult, it can be limiting for certain applications. For example, single photon detectors based on a competing technology, the transition-edge sensor (TES), are used successfully in microanalysis applications involving a scanning electron microscope (SEM).[4] An SEM is used to locally excite a sample of interest, while a TES detector measures the spectroscopic composition of the luminescent x-ray photons, allowing identification of the host material. A limitation in the TES performance comes from their relatively slow count rate, which is exceeded by the STJ detector by almost an order of magnitude. However, STJ detectors are not as feasible for this application, since the applied magnetic field deflects the electron beam, necessitating large sample-detector spacing. Removing the need for a magnetic field could thus open up new applications for the STJ detector. In addition, in scaling to larger arrays of STJ detectors, the magnetic field suppression of the supercurrent can become difficult. Small differences in junction fabrication may necessitate a slightly different field for each junction, requiring a separate electrical lead for each junction. Removing the need for a magnetic field is thus an attractive option.

In this paper we propose a new biasing scheme for an STJ detector based on a circuit of multiple junctions which removes the need for a magnetic field. The biasing occurs through the nonlinear interaction of the detector junction(s) with other junctions in the circuit. This nonlinear state can only be obtained with multiple junctions and not with resistors. The circuit is still compatible with high fidelity pulse readout, and the bias voltage is numerically stable against external fluctuations. Here we describe the circuit concept, show simulations to demonstrate its operation, and discuss practical considerations for detector designs.

The proposed circuit is shown in Fig. 1. The detector junction (junction 3) is placed in series with a second junction (junction 2) and then both are placed in parallel with a third junction (junction 1). The ratio of the critical currents is 1, 0.5 and 0.5 for junctions 1, 2 and 3 respectively; other combinations are possible. A current ($I_T$) is applied to the three junctions as shown. Fig. 2 shows how to reach desired operating state. First the current $I_T$ is increased until all three junctions have switched into the finite voltage state. Then the current is reduced to approximately the operating current shown in Fig. 2. Summing the voltages around the loop requires that $V_1 = V_2 + V_3$. In the desired state, junction 1 is biased at the superconducting energy gap ($V_1 = V_g = 2\Delta/e$) and junctions 2 and 3 are biased at half the energy gap ($V_2 = V_3 = V_g/2$). At this point junctions 2 and/or 3 can function as a detector; no magnetic field has been applied.[5]

We first show nonlinear simulations of the biasing state and then discuss practical considerations for detector design. To simulate the circuit, we solve the usual RSJ model[6] for each junction with an added term for the subgap current. The normalized current through junction $j$ is given by

$$i_j = h_j\left(\frac{d^2\varphi_j}{d\tau^2} + \Gamma(v_j)\frac{d\varphi_j}{d\tau} + \sin\varphi_j + i_{ss}(v_j)\right). \tag{1}$$

Here $\varphi$ is the phase difference across a given junction, $v$ is its normalized voltage ($v = d\varphi/d\tau$), $\tau = \omega_p t$ is the normalized time, $\omega_p$ is the plasma frequency, $t$ is time, $\Gamma(v)$ is the voltage-dependent damping, $h$ is the anisotropy parameter for the size of the different junctions, and $i_{ss}$ is the voltage-dependent subgap current. The subscript $j$ runs over the three junctions in the circuit. The currents $i_j$ are normalized to the critical current of the first junction, $I_{c1}$. The plasma frequency is given by $\omega_p = \sqrt{\frac{2\pi I_c}{\Phi_0 C}}$, where $\Phi_0$ is the flux quantum and $C$ is the junction capacitance. The BCS subgap current is given by:[7]

$$i_{ss} = \frac{2}{I_{c1}R_n e}\left(\frac{2\Delta}{eV + 2\Delta}\right)^{1/2}(eV + \Delta)e^{-\Delta/kT}\sinh\left(\frac{eV}{2k_BT}\right)K_0\left(\frac{eV}{2k_BT}\right), \tag{2}$$

where $k_B$ is Boltzman´s constant, $T$ is the temperature, and $K_0$ indicates the zero-order modified Bessel function. The nonlinear damping parameter $\Gamma$ is given by:

$$\Gamma = \Gamma_N g(v), \qquad (3)$$

where $\Gamma_N$ is the damping in the normal state, equal to $\left[\dfrac{\Phi_0}{2\pi I_C R_N^2 C}\right]^{1/2}$, and $g(v)$ is the voltage-dependent damping. To account for the damping in the subgap region and for the gap rise, we use the following empirical form for $g(v)$:[8]

$$g(v) = g_{sg} + \dfrac{1 - g_{sg}}{2}\left\{1 - \tanh\left[100\left(1 - \dfrac{v_j}{v_g}\right)\right]\right\}. \qquad (4)$$

Here $g_{sg}$ is a constant damping in the subgap region and $v_g$ is the normalized gap voltage. Equations (3) and (4) give $\Gamma(v) = g_{sg}\Gamma_N$ for $V < (2\Delta/e)$ and $\Gamma(v) = \Gamma_N$ for $V > (2\Delta/e)$.

To write the equations of the circuit we use fluxoid quantization and current conservation. Fluxoid quantization gives $(\varphi_1 - \varphi_2 - \varphi_3) = 2\pi f_{ind}$, where $f_{ind}$ is the induced frustration in the loop formed by the three junctions. Current conservation gives $i_1 = i/2 - i_m$ and $i_2 = i_3 = i/2 + i_m$; here $i_j$ is given by equation (1), $i = i_1 + i_2$, and $i_m$, the mesh current, is related with $f_{ind}$ through $i_m = 2\pi\lambda f_{ind}$. The parameter $\lambda = \Phi_0/(2\pi L I_{c1})$ measures the importance of induced fields; $L$ is the geometric inductance of the loop.[9]

The results of the model are shown in Fig. 2. The parameters used are $\Gamma_N = 0.02$, $h_1=1$, $h_2=0.5$, $h_3=0.5$, $g_{sg} = 10^{-4}$, and $\lambda=10$. Initially, as $I_T$ is increased, all three junctions are in the zero-voltage state. At a current of approximately $I_T = 1.3 I_{c1}$ the system switches to a running mode, where $V_1 = V_g$ and $V_2 = V_3 = (V_g/2)$. In real experiments the switching of the three junctions is affected by fluctuations, which are not included in Fig. 2. In any case, for reaching a detector biasing point, these switching dynamics are unimportant; one simply increases the current to a value large enough to ensure that all three junctions have switched. Once this point has been reached, the current is then decreased to the operating current shown, approximately $I_T = 0.2I_{c1}$. Operating currents as large as $I_T = 0.5I_{c1}$ give stable dynamics, but for noise purposes lower values of $I_{c1}$ are more desirable. Following this procedure ensures the desired state will be obtained. If the current is decreased further, the system will retrap to the zero voltage state across all three junctions. Fig. 2 plots the *total* current $I_T$ on the y-axis; at the operating point shown, nearly *all* of this current $I_T$ is flowing though junction 1. Only the small subgap current flows through junctions 2 and 3. This subgap current for two different temperatures is shown by the solid lines in Fig. 3a; they resemble a typical STJ I-V curve.

This state of the detector was motivated by a type of nonlinear dynamical state in a Josephson array called a breather, which is an intrinsically localized mode that exists in a ladder array driven by a uniform current.[10] The so-called type B breather has a similar voltage pattern of $V_g$, $V_g/2$ and $V_g/2$ for three junctions around a loop. Although not identical, the circuit in Fig. 1 has similar hysteretic dynamics to the type B breather state.

We have found that other nonlinear states coexist in the circuit with desired detector state, which we refer to as the symmetric state. The first is a state for which $V_1 = V_2 = V_g$ and $V_3 = 0$ (or $V_3 = V_g$ and $V_2 = 0$). We have also found a state for which $V_1 = V_g$, $V_2 = (V_g - \delta)$ and $V_3 = \delta$ (or $V_3 = [V_g - \delta]$ and $V_2 = \delta$), where $\delta \ll V_g$. These asymmetric states are undesirable. If the system starts out in the symmetric state, it can switch to one of the asymmetric states due to fluctuations. In order to check the stability of the symmetric state against external fluctuations, we added a noise term to each junction in equation (1) and then integrated the equations for the circuit. We chose a level of noise orders of magnitude larger than expected for thermal fluctuations at typical detector operating temperatures. Extrapolating our results to realistic amounts of noise, we find that the symmetric state is numerically stable for times orders of magnitude longer than typical experimental times ($10^4$ s).

The main fabrication requirement for reaching this biasing state is that the junctions be highly underdamped. Values of $\Gamma_N < 0.05$ result in the ideal dynamics that are shown in Figures 2 and 3. For $0.1 > \Gamma_N > 0.05$, the nonlinear biasing state still exists, but with an increased DC current through junctions 2 and 3, which will result in some excess noise. For $\Gamma_N > 0.1$, the dynamics becomes more complicated than we have shown in Figures 2 and 3. For niobium (Nb) junctions, achieving $\Gamma_N < 0.05$ requires fabricating a current density ($J_c$) of about 200 A/cm$^2$ or less, which is satisfied by most detector junctions tested to date. For aluminum (Al) junctions the current density should be lower, around $J_c \sim 5$ A/cm$^2$. This is achievable in most fabrication processes, although many Al junctions tested to date have values of $J_c$ slightly higher, around 30 A/cm$^2$. The loop inductance parameter, $\lambda$, appears to have no major constraints. The results shown are for $\lambda = 10$, but we have found the same results for $\lambda = 0.1$ and $\lambda = 1$.[11] We comment more on the possible circuit parameters in a future paper.[12]

With the circuit biased at the desired operating point, away from the hysteresis and switching in Fig. 2, a simpler dc model can be used to predict currents and voltages at the operating point of the circuit with the same accuracy as the full model. These results can be used to show the operation of the detector in response to an absorbed photon and to discuss issues of impedance and noise. In the dc model, we change equation (1) to $i_j = h_j \left[ \Gamma(v_j) v_j + i_{ss}(v_j) \right]$ and keep the relations $V_1 = V_2 + V_3$; $i_2 (V_2) = i_3$

($V_3$); and $i = i_1 (V_1) + i_2 (V_2)$, defining a system of algebraic equations which can be numerically solved. At the operating point $V_1 = V_g$ and $V_2 = (V_g - V_3)$. We can then solve the dc model equations graphically by plotting $I_3 = I_2$ versus $V_2$ and versus $(V_g - V_3)$. The intersection gives the operating point of the circuit, as shown in Fig. 3a.

To simulate the response to a photon, we increase the temperature of the detector junction. In a real detector the temperature first increases, as the excess quasiparticles tunnel through the junction, and then decreases, as the quasiparticles either recombine or diffuse away from the barrier. The detector junction(s) can be junction 2, junction 3, or both. The temperature of junction 1 stays constant. In Fig. 3a we show how the operating point can be followed as the subgap current increases. In the case of heating both junctions 2 and 3, the operating point moves straight up from A to B, and then back down from B to A. If only junction 3 is used then it moves to the left and up.

The current through junctions 2 and 3 as a function of temperature from the model is shown in Fig. 3b, for the case of heating both junctions. It is orders of magnitude less than $I_T$, showing again that only the small subgap current flows through junctions 2 and 3. Point A is for $(kT/\Delta) = (1/8)$, and point B is for $(kT/\Delta) = (1/9)$, corresponding to points A and B indicated in Fig. 3a. To read out the excess tunneling current, a current amplifier can be AC coupled and placed in parallel with junction 3, as shown in Fig. 1. (If both junctions are heated, a second amplifier can be used in parallel with junction 2.) The blocking capacitor, $C_B$, passes signal frequencies and block DC currents. Junctions 2 and 3 are high impedance, since they biased in the subgap region, making signal collection through the low impedance current amplifier relatively efficient.

The extra junctions in the circuit will add additional electronic noise to the amplifier which can potentially degrade the energy resolution of the detector. In many cases the additional noise will cause only a small, even negligible change in the total energy resolution. Under some conditions, the cost can be as large as a factor of two in the energy width. The two main sources of electronic noise are (i) the shot noise due to the current flowing through the junctions, and (ii) the voltage noise of the amplifier, which is converted to current noise due to the junction impedance.[13] The shot noise (i) is proportional to the DC current flowing through each junction. The current flowing through junctions 2 and 3 is the same subgap current that usually flows through a single junction; hence there is no increase over the usual shot noise. The extra current flowing in junction 1 will give some additional shot noise; however, since the impedance of junction 1, biased at the energy gap, is many orders of magnitude smaller than the impedance of junctions 2 and 3, this extra noise current will almost exclusively flow through junction 1 and not through junctions 2 and 3. Thus it will not add significantly to the total shot noise seen by the amplifier. For the voltage noise (ii), the new impedance seen by the amplifier will be approximately junction 2 in parallel with junction 3. If junctions 2 and 3 have the same dynamic resistance, then the new impedance seen by the amplifier will be a factor of two smaller than the impedance of junction 2 or 3 individually. This increases the contribution of the amplifier voltage noise to the total current noise by a factor of two as compared to a single junction. The actual amount of increase in the energy width caused by the increases in electronic noise will depend on the particular detector geometry, photon energy, and junction size.

In addition to noise there are other possible considerations for detector design. In our models we have assumed the theoretical BCS subgap current ($i_{ss}$) along with an additional linear subgap impedance ($g_{sg}$). In real devices there are often other contributions to the subgap current, including Fiske modes, normal metal tunneling, and multi-particle Andreev reflection. The importance of these issues will need to be determined through experiments and junction fabrication. A few points should be noted, however. The first is that one need not be constrained by the ratio of 1.0, 0.5 and 0.5 for the critical currents of the three junctions. We have seen that several different ratios give subgap biasing, such as 1.0, 0.5, and 0.3. Different ratios will result in different operating voltages in the subgap region, which can be chosen to avoid any known subgap features. In addition, by using the technique we have described, junction shapes will not be as constrained as they are presently. In existing devices, junction shapes are usually "stretched" in order to allow more efficient supercurrent suppression by the external magnetic field. Since the external field is no longer needed, the junctions can be chosen to be any shape. This would help, for example, in avoiding Fiske modes.[14] Finally, since two of the junctions in the circuit can each be used as a detector, there are possibilities to bias a two-junction detector or even a many-pixel array in the future with this technique. This will free up electrical leads and offer more flexibility in overall experiment design.

We thank D.E. Prober, M.C. Gaidis, S. Friedrich and L. Frunzio for useful discussions. J.J.M. acknowledges the FEDER program through Project No. BMF2002-00113. This project was partially supported by the NSF grant DMR-9988832. K.S. acknowledges financial support from Colgate University and from the Colgate Division of Natural Sciences.

**Figure Captions:**

Fig. 1: Circuit schematic for magnetic-field free detector biasing. The X's represent junctions. A current $I_T$ is applied and splits between the two branches, with the same current flowing through junctions 2 and 3. The pulse amplifier is AC coupled to junction 3. The excess quasiparticles due to photon absorption can be read out through junctions 2, 3 or both, depending on where the photon is absorbed.

Fig. 2: Hysteretic dynamics for the circuit shown in Fig. 1. The total current $I_T$ is plotted against the voltage for junctions 1 (dotted line), 2 and 3 (both solid lines). $I_T$ is initially increased until the three junctions switch to non-zero voltage; then the current is decreased. At the operating current, junctions 2 and 3 are in the subgap region where they can function as photon detectors; here nearly all the current $I_T$ is flowing through junction 1.

Fig 3: Detector operation in the subgap region. (a) I-V curves showing the subgap current for junctions 2 and 3. $I_2 = I_3$ is plotted versus $V_3$, solid line, and ($V_g$-$V_3$), dotted line. The two lower curves are for a temperature $(kT/\Delta) = (1/9)$ and the two upper curves are for $(kT/\Delta) = (1/8)$. At the lower temperature the intersection of the two curves gives the operating point. The head from a photon, coupled to both junctions, causes the system to move from A to B, straight up. If the head were only coupled to one of the junctions it would move to the left as shown. (b) Current through junctions 2 and 3 as a function of reduced temperature. The two end points indicate the change in moving from A to B as in Fig. 3a. The system will heat up from A to B and then cool back down from B to A.

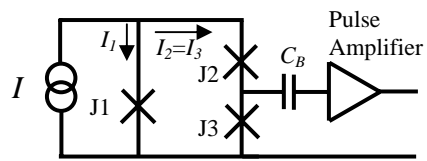

Fig. 1

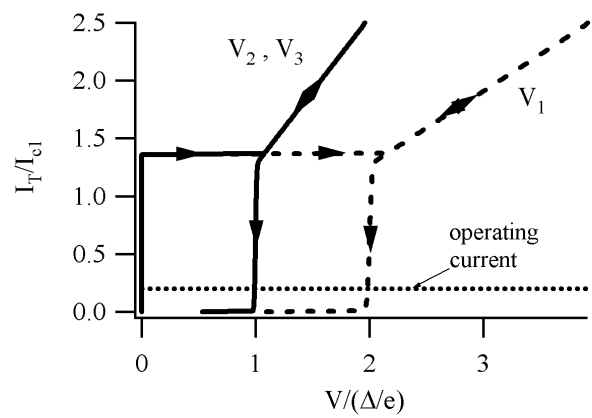

Fig. 2

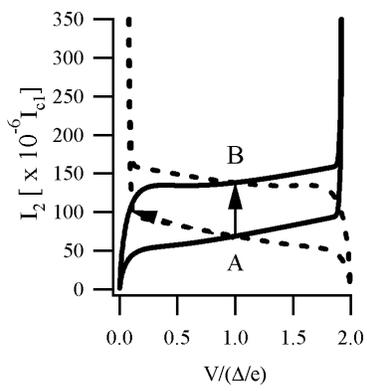 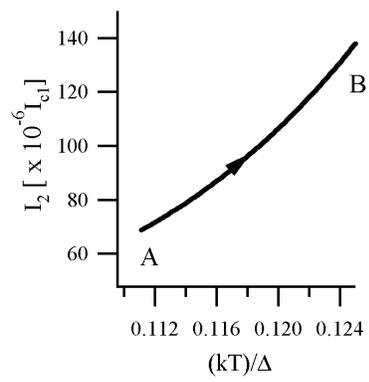

(a) (b)

Fig. 3